# A comparison of phase change materials in reconfigurable silicon photonic directional couplers

Ting Yu Teo,[1,4] Milos Krbal,[2] Jan Mistrik, [2,3] Jan Prikryl,[2] Li Lu,[1] and Robert Edward Simpson[1,5]

[1] Singapore University of Technology and Design, 8 Somapah Road, 487372 Singapore, Singapore
[2] University of Pardubice, Faculty of Chemical Technology, Center of Nanomaterials and Nanotechnologies (CEMNAT), Legions Square 565, 530 02 Pardubice, Czech Republic
[3] University of Pardubice, Faculty of Chemical Technology, Institute of Applied Physics and Mathematics, Studentska 95, 532 10 Pardubice, Czech Republic
[4] tingyu_teo@mymail.sutd.edu.sg
[5] robert_simpson@sutd.edu.sg

**Abstract**

The unique optical properties of phase change materials (PCMs) can be exploited to develop efficient reconfigurable photonic devices. Here, we design, model, and compare the performance of programmable $\mathbf{1 \times 2}$ optical couplers based on: $Ge_2Sb_2Te_5$, $Ge_2Sb_2Se_4Te_1$, $Sb_2Se_3$, and $Sb_2S_3$ PCMs. Once programmed, these devices are passive, which can reduce the overall energy consumed compared to thermo-optic or electro-optic reconfigurable devices. Of all the PCMs studied, our ellipsometry refractive index measurements show that $Sb_2S_3$ has the lowest absorption in the telecommunications wavelength band. Moreover, **$Sb_2S_3$** -based couplers show the best overall performance, with the lowest insertion losses in both the amorphous and crystalline states. We show that by growth crystallization tuning at least four different coupling ratios can be reliably programmed into the $Sb_2S_3$ directional couplers. We used this effect to design a 2-bit tuneable $Sb_2S_3$ directional coupler with a dynamic range close to 32 dB. The bit-depth of the coupler appears to be limited by the crystallization stochasticity.

## 1. Introduction

Reconfigurable photonic devices have the potential of revolutionizing Photonic Integrated Circuits (PICs) [1]. Currently, most PICs are custom- made for a specific purpose, which is analogous to the electrical Application Specific Integrated Circuits (ASICs) technology. Developing reconfigurable photonic devices will introduce greater flexibility to photonics engineers. Thus, developing programmable photonics is critical in advancing PIC technology. Just as Field Programmable Gate Arrays (FPGAs) paved the way for modern electrical integrated circuits technology, such as hardware neural networks, we believe programmable PICs will pave the way for innovative photonics devices, such as practical implementations of all-optical neural network chips.

Mach Zehnder Interferometers (MZIs) are one of the basic building blocks of programmable PICs. [2] These devices consist of optical directional coupler switches that split incoming light into two coherent waves, and the signal is programmed by controlling the relative phase difference and intensity in the two waveguide interferometer arms. Common tuneable directional couplers make use of thermo-optic or electro-optic effects to tune the refractive index [3-5]. However, they require a continuous power supply to maintain their optical properties. This is non-ideal for systems that require highly interconnected waveguide meshes or networks. For example, in optical neural networks, each reprogrammable synaptic weight that makes use of the thermo-optic or electro-optic effect require an additional 10 mW [6]. The network becomes inefficient when processing large amounts of data, and ultimately the power requirement limits the scale of the network.

Incorporating chalcogenide phase change materials (PCMs) into directional couplers to introduce tuneability is desirable because the PCM does not require power to hold its optical state. Moreover, many PCMs can reversibly switch between their amorphous and crystalline states on a sub-nanosecond time scale [7, 8]. These properties provide a means to design sophisticated tuneable directional couplers [9-11].

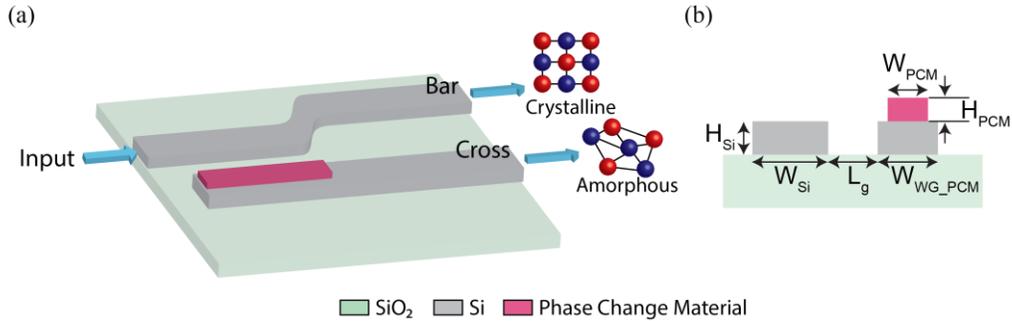

**Fig. 1.** PCM- tuned directional coupler design. (a) PCM directional coupler switching operation. In the amorphous state the output signal is at the cross port due to evanescent coupling. In the crystalline state, the signal stays in the same waveguide due to phase mismatch between the two waveguides. (b) Cross section of the PCM directional coupler. The labelled dimensions are critical when designing the couplers.

A **1 × 2** PCM directional coupler was designed that consists of two output waveguide ports, with one waveguide containing a layer of PCM deposited as shown in Fig. 1a. Fig. 1b shows the parameters to be determined when designing PCM directional couplers. The PCM directional coupler is designed to evanescently couple light from a Si waveguide into the PCM -integrated waveguide port (cross) in the amorphous state. Crystallizing the PCM decreases the amplitude of the evanescent field, and the light remains in the input Si waveguide (bar) port. This design exploits the large refractive index change between the two PCM structural states to control the phase matching condition between the waveguides.

Previous works [9-12] have demonstrated the reconfigurable functionality in PCM directional couplers. In those designs, $Ge_2Sb_2Te_5$ and $Ge_2Sb_2Se_4Te_1$ have been used because they display a large change in refractive index, **Δ$n$**, upon phase transition. The higher loss in the crystalline state is avoided as the light remains in the bar port. However, they still exhibit non-negligible losses at 1550 nm in the amorphous state [13]. To avoid these high optical losses, we propose the use of low-loss PCMs, $Sb_2S_3$ and $Sb_2Se_3$. Previous works [14-16] showed that both materials exhibit distinguishable optical constants upon a structural phase transition and have a lower extinction coefficient than $Ge_2Sb_2Te_5$ and $Ge_2Sb_2Se_4Te_1$ in both the amorphous and crystalline telecommunications spectral band.

In this work, we aimed to assess the suitability of using low-loss PCMs to program the coupling ratio of directional couplers. This was done by designing, modelling, and comparing the performance of programmable **1 × 2** optical couplers based on: $Ge_2Sb_2Te_5$, $Ge_2Sb_2Se_4Te_1$, $Sb_2Se_3$, and $Sb_2S_3$ PCMs. Using three- dimensional Finite Difference Time Domain (3D FDTD) calculations, we showed that at the telecommunication wavelength, all PCM directional couplers exhibited low insertion losses <−1.5 dB and crosstalk of between – 20 dB to – 40 dB in both the amorphous and crystalline states. In particular, $Sb_2S_3$ – tuned directional couplers exhibited the lowest insertion loss in both structural states. This corroborates with our optical constant measurements, with Sb$_2$S$_3$ having the lowest extinction coefficient, $\boldsymbol{k}$. We then studied how crystallization proceeds along strips of Sb$_2$S$_3$, and the gradual effect of crystallization on the device transmission. Narrow strips of Sb$_2$S$_3$, similar to those patterned on the directional coupler, were annealed above the glass transition temperature. We observed that unlike Ge$_2$Sb$_2$Te$_5$, Sb$_2$S$_3$ crystallization is growth- dominated [17] and the crystal growth tends to proceed progressively along the strip. This suggests that the crystallized length can be controlled with temperature. Hence, multiple coupling ratios can be achieved through partial crystallization. However, we observe that the growth-dominated crystallization process is stochastic, which makes it challenging to introduce more than four multi-level states. The stochasticity can be minimized by optimizing the programming method.

## 2. Low-loss PCM Directional Coupler Design and Modelling

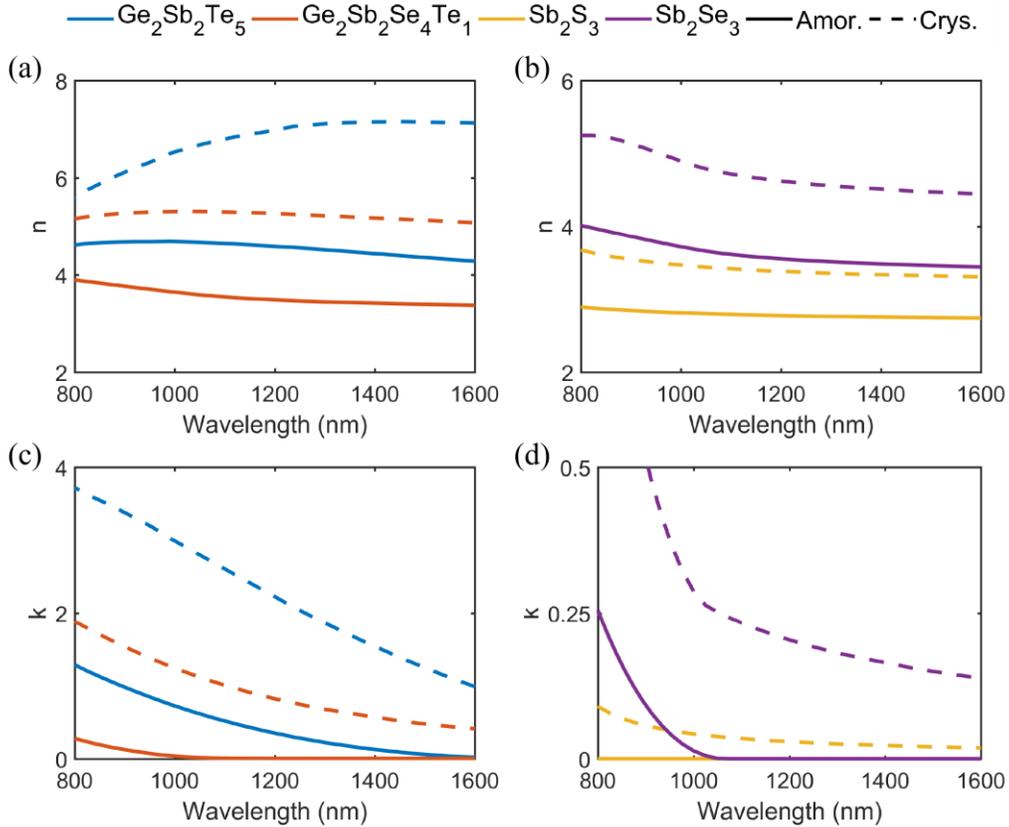

**Fig. 2.** Optical properties of the four PCMs. Refractive index, $\boldsymbol{n}$, of (a) GST and GSST, (b) Sb$_2$S$_3$ and Sb$_2$Se$_3$. Extinction coefficient, $\boldsymbol{k}$, of (c) GST and GSST, (d) Sb$_2$S$_3$ and Sb$_2$Se$_3$ in their respective amorphous and crystalline states.

To design PCM-tuned couplers, we considered four different potentially useful PCMs: Ge$_2$Sb$_2$Te$_5$, Ge$_2$Sb$_2$Se$_4$Te$_1$, Sb$_2$Se$_3$, and Sb$_2$S$_3$. Fig. 2 shows the optical constants of the four materials. The values of Ge$_2$Sb$_2$Te$_5$ and Ge$_2$Sb$_2$Se$_4$Te$_1$ were extracted from references [10, 18] whilst the optical constants of Sb$_2$S$_3$ and Sb$_2$Se$_3$ were measured by Variable Angle Spectroscopic Ellipsometry (VASE). The Sb$_2$S$_3$ and Sb$_2$Se$_3$ thin films used in the ellipsometry measurements were prepared using Radio frequency (RF) magnetron sputtering and pulsed-laser deposition (PLD), respectively. Both depositions were performed without heating the substrate. The films were 160 nm and 200 nm thick, respectively. To crystallize the $Sb_2S_3$ and $Sb_2Se_3$ films, the samples were annealed to 320°C and 250 °C, respectively. Further details of the deposition parameters, the annealing process, and the ellipsometry measurement procedure can be found in the Supplemental Document sections 1 and 2.

Determining the optical constants of the PCM becomes critical when modelling PCM-tuned directional couplers. This is because the PCM layer controls the amplitude of the evanescent field between the two waveguides. In the amorphous state, the two waveguides must have similar refractive indices for the optical signal to couple strongly into the cross port. Hence, the absolute values of the refractive index, **n**, are important. This is unlike other single-waveguide reconfigurable photonic devices, such as optical phase shifters and ring resonators, where only a *change* in PCM refractive index is needed to induce a relative phase or resonance shift upon a structural phase transition [19-21]. However, the accuracy of ellipsometry measurements of refractive indices is mainly limited by the dispersion model. I.e. the dispersion model choice will affect the refractive index measurement and concomitantly affect the modelled coupler performance

The band structure of semiconductor materials should be considered when fitting optical constants to ellipsometry data. Hence, in our ellipsometry fitting, we considered dispersion models that describe the material optical absorption above and below the material's bandgap. Moreover, the imaginary and real components of the dielectric function are intwined by causality, hence any physically realistic model to describe the optical properties of a semiconductor should be Kramers–Kronig consistent. For this reason, the Tauc-Lorentz model was used to fit the $Sb_2S_3$ and $Sb_2Se_3$ ellipsometry measurements in this work.

The $Sb_2Se_3$ and $Sb_2S_3$ band gaps derived from the Tauc-Lorentz dispersion model to the ellipsometric constants were comparable to those reported in the literature. In the amorphous state, the Tauc-Lorentz model showed bandgap energies of 2.05 eV and 1.16 eV for $Sb_2S_3$ and $Sb_2Se_3$, respectively. These values are consistent with those reported in references **[15, 22]**. Upon crystallization, the materials became polycrystalline and a single Tauc-Lorentz model was no longer sufficient to represent the optical constants over a broad spectral range. Hence, additional Lorentz oscillators were necessary to fit the optical constants of the polycrystalline material. These additional oscillators relate to critical points of the electronic band transition. Both materials required four Lorentz oscillator models to supplement the basic Tauc-Lorentz model. For both materials, the number and position of the critical points are comparable to previous works **[23, 24]**. Moreover, for $Sb_2Se_3$, the electronic band transition energies at these critical points are comparable to those used in other dispersion models, and this adds a degree of confidence to our ellipsometry fitting results **[23]**. Upon fitting the ellipsometry measurements with the Tauc-Lorentz dispersion model, we obtained the optical constants of the amorphous and crystalline low-loss PCMs shown in Figs. 2b and 2d. Overall, in the near-infrared spectrum, the low-loss PCMs have a lower extinction coefficient, **k,** and lower change in refractive index, **Δn**, when compared to the Telluride-based materials. Both low-loss PCMs exhibit negligible losses (close to zero) in the amorphous state. Of all the PCMs in the crystalline state studied, $Sb_2S_3$ has the lowest extinction coefficient, see Fig. 2d, which is likely due to its wider bandgap.

As a rule, there is a decrease in **k** when one moves up the Chalcogen group of the periodic table (from Te to Se to S) due to the optical band gap opening [25] and we also see that the extinction coefficient in Fig. 2 exhibit a similar trend. A wavelength of 1550 nm corresponds to a photon energy of 0.8 eV and the bandgap of amorphous $Ge_2Sb_2Te_5$ is 0.7 eV in the amorphous state and 0.5 eV in the face-centered cubic state [26]. Hence, 1550 nm light is absorbed by interband transitions in $Ge_2Sb_2Te_5$. In contrast, the bandgap of $Sb_2S_3$ is 2.05 eV in the amorphous state and 1.7 eV in the crystalline state [15]. Thus, it has the smallest $k$ due to its largest optical bandgap. Despite its large bandgap, $Sb_2S_3$ can be switched by either laser heating with above-bandgap light, heating with below bandgap light and using a resonant structure, or using electrical Joule heat-induced phase transitions [15]. Unlike crystalline $Ge_2Sb_2Te_5$, $Sb_2S_3$ was recently shown to exhibit birefringence [16]. This difference stems from the $Sb_2S_3$ crystal domains being larger than the wavelength of light, whereas the domains in $Ge_2Sb_2Te_5$ are below the diffraction limit, and therefore the polycrystalline film exhibits an isotropic refractive index. Amorphous marks that were smaller than the diffraction limit could also be written into $Sb_2S_3$ thin films using femtosecond pulses at 780 nm [16]. This result, together with the measured low $k$ value, suggest that $Sb_2S_3$ might be suitable for designing multi-bit programmable couplers.

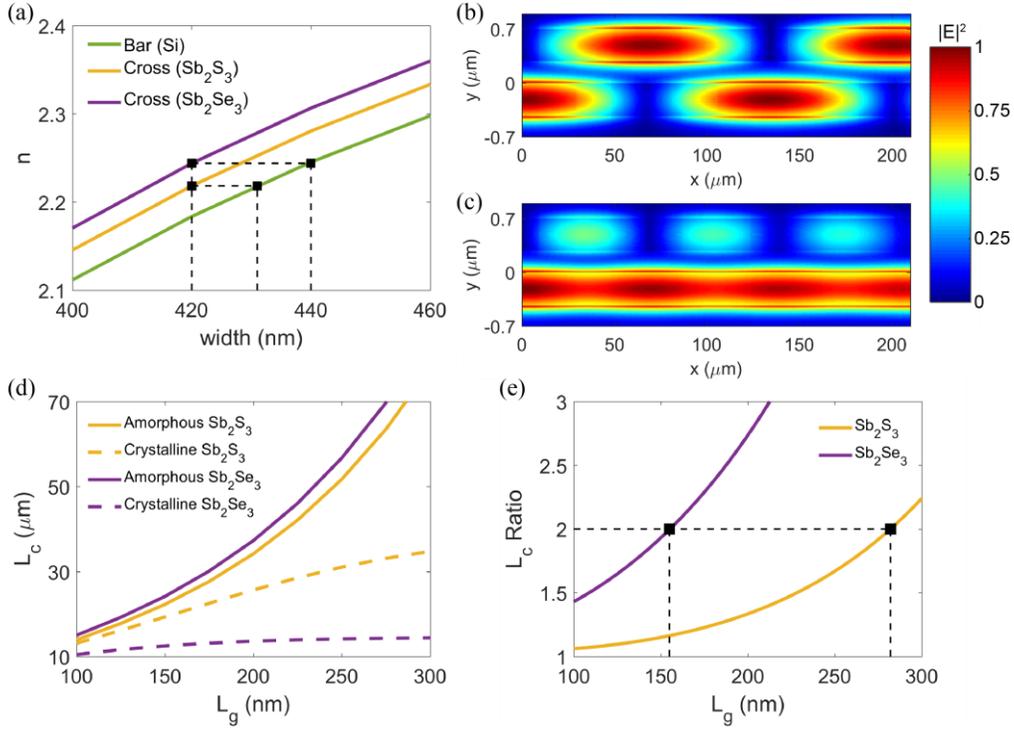

**Fig. 3.** PCM waveguide design process. (a) Waveguide width optimization for phase matching between bar and corresponding cross waveguides. Optical field intensity propagation for (b) amorphous and (c) crystalline $Sb_2S_3$ when $L_{c,amor} \approx 2m \cdot L_{c,crys}$, where $m \in \mathbb{Z}^+$. The optical signal propagates through different waveguide ports depending on the PCM state, consistent to current PCM directional coupler switching behavior. (d) Coupling length of amorphous and crystalline $Sb_2S_3$ and $Sb_2Se_3$, and (c) Coupling length ratio of $Sb_2S_3$ and $Sb_2Se_3$.

Using the optical constants in Fig. 2, we modelled $Ge_2Sb_2Te_5$, $Ge_2Sb_2Se_4Te_1$, $Sb_2Se_3$, and $Sb_2S_3$ Silicon-waveguide directional couplers. The devices were optimized in the transverse electric (TE) mode. Silicon-based couplers were chosen due to their wide applicability in integrated photonics and integrability with electronic integrated circuits. When designing PCM directional couplers, the dimensions shown in Fig. 1b was determined. They were the (a) dimensions of the phase change material, cross, and bar waveguide, (b) coupling gap between the two waveguides, $L_g$, and (c) the coupling length, $L_c$. $L_c$ is equivalent to the length of the phase change material.

To ensure fair comparison across the four PCM-tuned directional couplers, we fixed the PCM-tuned waveguide dimensions. The width of the PCM – tuned waveguide, represented as $W_{WG\_PCM}$ in Fig. 1b, is fixed at 420 nm and the PCM width and thickness were set at 320 nm and 20 nm respectively. These dimensions were adopted from reference [9] as they ensured single mode operation in $Ge_2Sb_2Te_5$ -tuned directional couplers. All waveguide heights were fixed at 220 nm. The widths of the bar waveguides were then optimized accordingly to satisfy the phase matching condition in the amorphous state [9]. Fig. 3a shows the waveguide dimensions required to attain phase matching conditions for $Sb_2S_3$ and $Sb_2Se_3$. The effective refractive indices of the cross and bar waveguides, $n_{eff}$, were calculated using the MODE solution waveguide solver in Lumerical [27]. The electric field distribution profiles of the PCM waveguides with the optimized dimension and the corresponding $n_{eff}$ values can be found in the supplementary document Fig. S3 and Table S1 respectively. The choice of $L_g$ was critical when designing the low-loss directional couplers. This is because the $\Delta n$ at 1550 nm between the crystalline and amorphous states is smaller than $Ge_2Sb_2Te_5$ and $Ge_2Sb_2Se_4Te_1$. For $Ge_2Sb_2Te_5$ and $Ge_2Sb_2Se_4Te_1$, the larger $\Delta n$ upon phase transition caused a large phase mismatch between the two waveguides in the crystalline state. Hence, the input signal did not couple into the cross-waveguide in the crystalline state. $L_g$ was thus chosen to optimize the trade-off between insertion loss in the crystalline state and a longer device coupling length [9]. However, for PCMs with a smaller $\Delta n$, the phase mismatch in the crystalline state was not as large and the coupling gap needs to be optimized to allow the coupling length in the amorphous state to be a common multiple of the coupling length in the crystalline state. This allowed the optical signal to couple back into the bar-waveguide in the fully crystallized state. Lumerical's Eigenmode Expansion solver (EME) [27] was used to demonstrate that this requirement is necessary. For clarity and analysis, the field propagation patterns in the amorphous and crystalline state of an $Sb_2S_3$ directional coupler are given in Figs. 3b and 3c, respectively. Since the effective coupling length of the directional coupler must allow the signal to exit at the different ports upon phase transition, the coupling length of each structural state must be an even multiple from each other: $L_{c,amor} \approx 2m \cdot L_{c,crys}$ where $m \in \mathbb{Z}^+$. To minimize the footprint and coupling losses, we chose the lowest common even multiple, where $m = 1$. The coupling length in the amorphous state, $L_{c,amor}$, was calculated using the coupled wave theory equation [9, 28]:

$$L_{c,amor} = \frac{\lambda}{2(n_1 - n_2)} \tag{1}$$

where $\lambda$ is the wavelength, and $n_1$ and $n_2$ are the effective refractive indices of the odd and even supermodes in a two-waveguide system. In the crystalline state, the change in the effective refractive index of the PCM-tuned waveguide results in a phase mismatch. To account for the phase mismatch, the crystalline coupling length was derived to be [29]:

$$L_{c,crys} = \frac{L_{c,amor}}{\sqrt{\left(\frac{\Delta\beta * L_{c,amor}}{\pi}\right)^2 + 1}} \tag{2}$$

Where $\Delta\beta$ is the effective propagation constant difference and is mathematically represented as $\Delta\beta = \frac{\Delta n_{eff}}{(2\pi/\lambda)}$. Again, the values of $n_1$, $n_2$ and $\Delta n_{eff}$ were calculated using the MODE solution solver within Lumerical. The electric field distribution patterns and the calculated values of the two-waveguide system can be found in Fig. S3 and Tables S2a and S2b respectively.

Upon stipulating the design requirements and calculations, the coupling gap, $L_g$, and length, $L_c$, of the Sb$_2$S$_3$ and Sb$_2$Se$_3$ directional couplers were then determined according to Figs. 3d and 3e. Fig. 3d shows the corresponding $L_c$ of the materials for the different structural state as $L_g$ increases. For clarity, we represent the $L_c$ of the two structural states as a ratio i.e. $L_c$ ratio $= \frac{L_{c,amor}}{L_{c,crys}}$ in Fig. 3e. $L_g$ was chosen to give a $L_c$ ratio of two. The final dimensions of the PCM directional couplers are given in Table 1. For comparison purposes, the Ge$_2$Sb$_2$Te$_5$ and Ge$_2$Sb$_2$Se$_4$Te$_1$ directional couplers were also modelled based on the design process reported in [9].

**Table 1.** Device dimensions and switching time [14, 15, 30, 31] for the various PCM-tuned directional couplers

| Material | PCM Width (nm) | PCM Height (nm) | Si Waveguide Width (nm) | Coupling Gap (nm) | Coupling Length (μm) | Switching time (ns), thickness (nm) |
|---|---|---|---|---|---|---|
| Ge$_2$Sb$_2$Te$_5$ | | | 456 | 107 | 18.40 | 50, 60 |
| Ge$_2$Sb$_2$Se$_4$Te$_1$ | | | 439 | 200 | 36.945 | 100*10$^6$, 40 |
| Sb$_2$S$_3$ | 320 | 20 | 431 | 282 | 67.175 | 78, 20 |
| Sb$_2$Se$_3$ | | | 440 | 155 | 25.277 | 100*10$^6$, 50 |

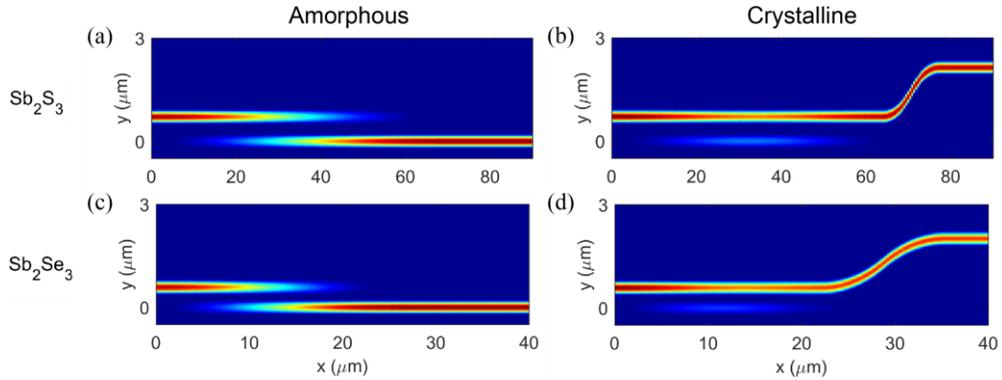

**Fig. 4.** Normalized power field distribution (arbitrary units) of (a) amorphous and (b) crystalline Sb$_2$S$_3$ and (c) amorphous and (d) crystalline Sb$_2$Se$_3$ directional coupler

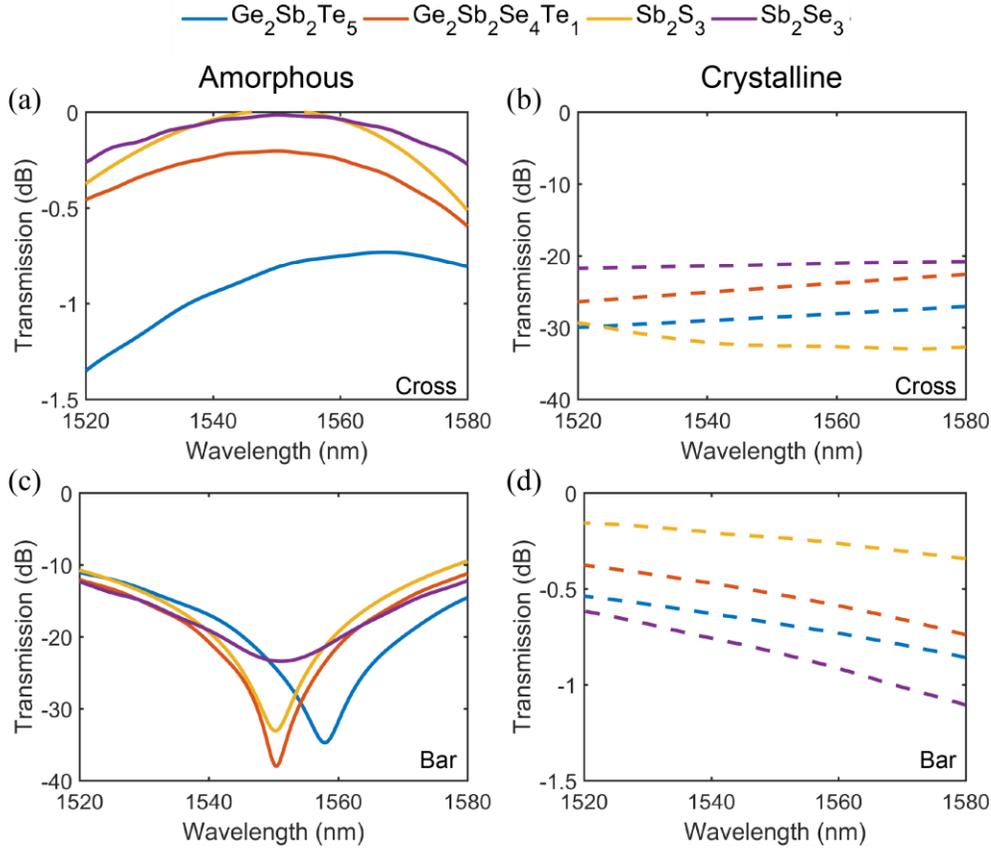

**Fig. 5.** PCM Directional coupler performance. The insertion loss of the coupler in the (a) amorphous and (d) crystalline state and crosstalk in the (b) crystalline and (c) amorphous state.

The insertion loss and cross talk for all four PCM couplers were analyzed in the telecommunication conventional band (1520 nm to 1580 nm) using a finite difference time domain (FDTD) approach to solve Maxwell's equations. Insertion loss represents the proportion of light that is transmitted through the coupler while the crosstalk represents the proportion of light in the inactive output port. The power field distribution and directional coupler performance are shown in Figs. 4 and 5 respectively. Figs. 4a to 4d show the power distribution of the low-loss coupler devices. The corresponding insertion loss and crosstalk of the four PCM devices, from a wavelength of 1520 nm to 1580 nm, are presented in Figs. 5a to 5d. All the PCM directional couplers studied had an insertion loss below -1.5 dB and a minimum crosstalk between −20 dB to −40 dB in both structural states. The switching characteristic of the $Sb_2S_3$ and $Sb_2Se_3$ directional couplers is consistent with previous works [9, 10], which adds a degree of confidence to the analysis of the coupler designs.

The $Sb_2S_3$ and $Sb_2Se_3$ programmable couplers have a higher transmission than the Telluride-based devices when the PCM is completely amorphous. From Fig. 5a, we observe that the insertion loss of these materials are low. At around 1550 nm wavelength range, both materials exhibit very low (close to 0 dB) insertion losses.

Upon crystallization, the $Sb_2S_3$ device still shows the lowest insertion loss. However, the insertion loss of the $Sb_2Se_3$-tuned devices becomes greater than the Telluride- devices. The results are illustrated in Fig. 5d. The higher insertion losses in $Sb_2Se_3$ are attributed to the design rule being $L_{c,amor} \approx 2 \cdot L_{c,crys}$. Hence, the signal in the $Sb_2Se_3$ bar port must couple into the PCM- waveguide and then back to the Si waveguide in the crystalline state. This requires the coupling losses and the extinction coefficient, ***k***, of the PCM material to be low. In contrast, the crystalline $Ge_2Sb_2Te_5$ and $Ge_2Sb_2Se_4Te_1$ exhibit a large phase mismatch between the cross and bar ports and the signal effectively remains in the bar port, without coupling. Thus, the transmissivities for the telluride PCMs in this coupler design are higher than $Sb_2Se_3$. However, the extinction coefficient, ***k,*** of $Sb_2S_3$ is sufficiently low that the insertion losses are still less than the Telluride- based directional couplers.

$Sb_2S_3$ has the best overall performance for programmable low-loss optical couplers. This is readily seen in Fig 6, which compares the performance of the different PCM couplers using a spider diagram. Importantly, $Sb_2S_3$ displayed the lowest insertion loss in both the amorphous and crystalline states and had the lowest crosstalk in the crystalline state. However, the better performance comes at the expense of a substantially longer coupling length, which is due to the smaller $Sb_2S_3$ $\Delta n$ upon phase transition.

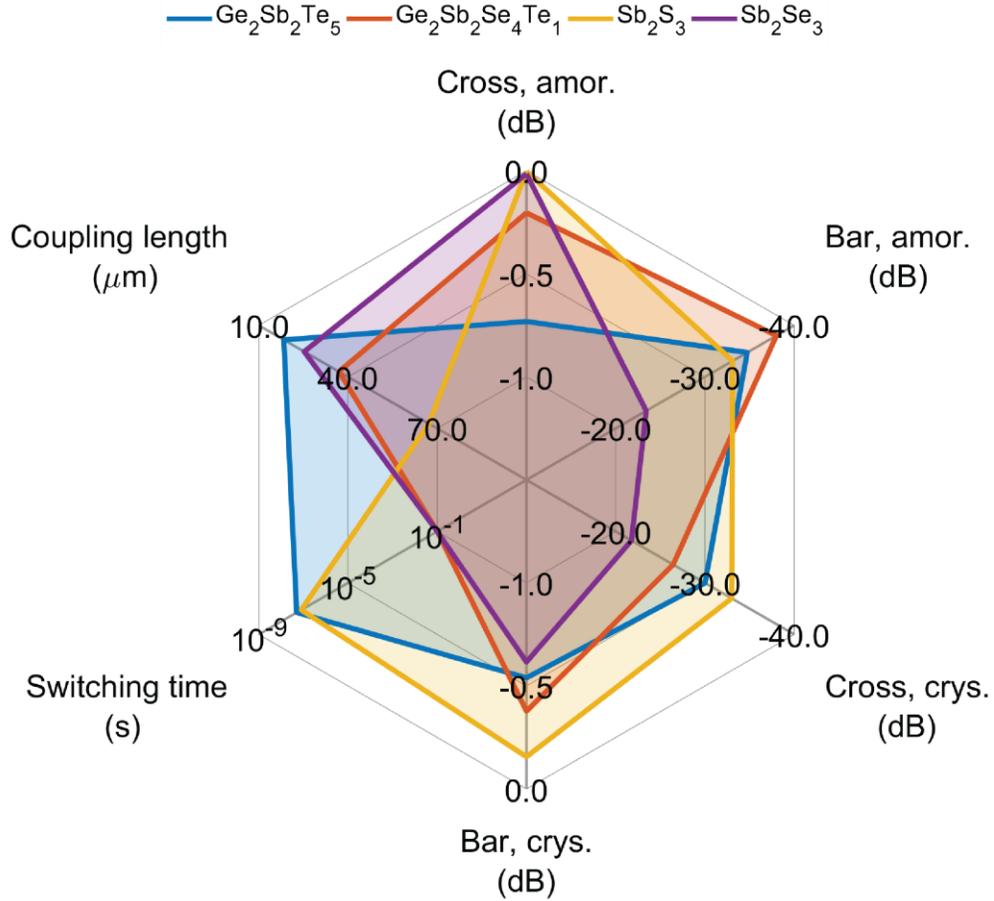

**Fig. 6.** Radar chart that compares the performance of the four PCM directional couplers. The switching times for each material are obtained from [14, 15, 30, 31]. The axes are arranged such that the desirable properties are on the edge of the chart.

With $Sb_2S_3$-tuned optical couplers having the better overall performance, these devices should be considered for non-volatile routing of optical signals through PICs that operate at the telecommunications wavelength. Here, we study how this material and coupler platform can introduce additional functionalities. One important area of study is partial crystallization of the PCM strip. Through studying the crystallization behavior of $Sb_2S_3$, we can implement multiple switching states and understand how the devices can be reliably switched into these different coupling ratios.

Intermediate switching states are a desirable feature of reprogrammable couplers. Indeed, partial crystallization of the PCM layer can be used to introduce multi-level optical states into a photonics device [32-35]. When crystallized to varying extents, the optical constant of the material changes [32]. There are essentially two ways to exploit crystallization to create a multilevel optical device and these depend on how the material crystallizes. Crystallization of some materials is dominated by nucleation. For example, $Ge_2Sb_2Te_5$ exhibits a high density of crystal nuclei when heated to intermediate temperatures above its crystallization temperature. This means that the crystallites tend to be small, and they grow in a short time [32]. Other materials, such as AIST and $Sb_2S_3$, do not easily nucleate and the crystallites grow large from very few nucleation centers. $Sb_2S_3$ does not easily nucleate, hence large crystallites can be observed under an optical microscope [16]. If the crystallites are substantially smaller than the wavelength of light, then the effective refractive index of the partially crystallized material can be derived with the Clausius-Mossoti relation [36, 37]. On the other hand, for $Sb_2S_3$, where the crystallites tend to be larger than the wavelength of light, it is more appropriate to compute the multiple transmission levels by solving Maxwell's equations for reflection and transmission of light propagating through the two material domains.

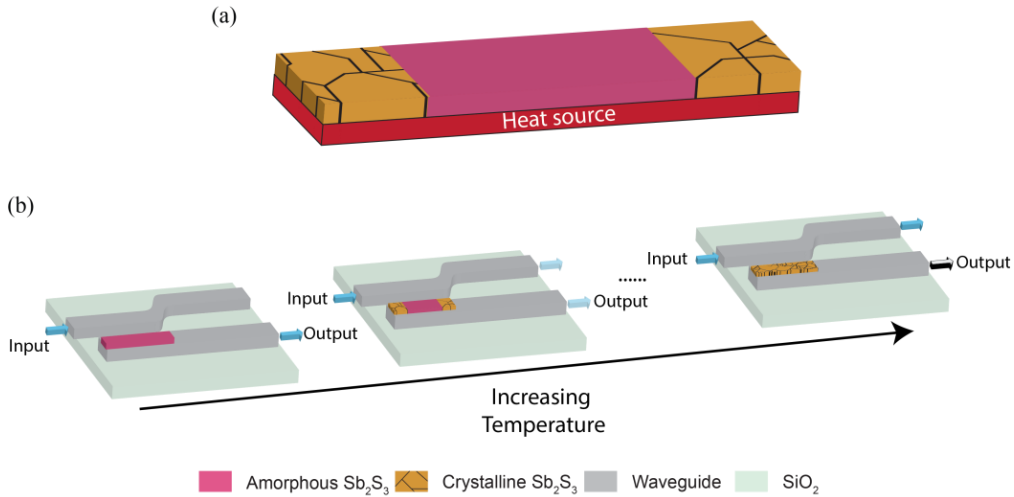

**Fig. 7.** Schematic of Growth Crystallization Tuning (GCT) mechanism (a) Imbedded heat source in the system to crystallize PCM (b) Tunable coupling ratio due to PCM crystallizing to different extent. The coupling ratio is controlled by the temperature of heat source as represented by a decreasing output intensity signal.

We envisage multiple ways to crystallize the PCM couplers. The first way is by using a laser to heat the PCM and create crystalline pixels along the waveguide length. However, this might be impractical for a small-scale device. A more practical design might include an embedded heater below the PCM [19, 20, 38]. Since $Sb_2S_3$ is dominated by crystal growth, this heater could be used to controllably crystallize strips of $Sb_2S_3$ to different extents, which in turn controls the coupling ratio, see schematic shown in Fig. 7. We name this method Growth Crystallization Tuning (GCT).

To exploit GCT in Sb$_2$S$_3$-tuned Si waveguide directional couplers, we first must understand how the Sb$_2$S$_3$ strips crystallize. Therefore, we patterned Sb$_2$S$_3$ cuboids of dimension 46.7 $\mu$m by 0.4 $\mu$m by 40 nm (Length by Width by Height) on a Silica on Si substrate. These dimensions are similar to those of the optimized coupler design. The strips were fabricated in the following sequence: (a) Electron Beam Lithography patterning, (b) Material deposition with Radio Frequency (RF) Magnetron Sputtering and (c) Material lift-off. We first deposited an 89 nm PMMA 950K A2 photoresist onto the substrate by spin coating at 2000 RPM for 60 seconds. Electron beam lithography was then performed using the Raith eLINE Plus lithography system to pattern the Sb$_2$S$_3$ strips. The electron acceleration voltage was 30 kV with an aperture size of 30 $\mu$m. The photoresist was then developed using the MIBK: IPA developer of ratio 1:3. Subsequently, we deposited Sb$_2$S$_3$ on the resist pattern using a 50.8 mm diameter 99.9% pure Sb$_2$S$_3$ target using the AJA Orion 5 sputtering system with a base pressure of 2.3×10$^{-7}$ Torr. The sputtering process took place in an Argon environment at a pressure of 3.7×10$^{-3}$ Torr. The RF power was set to 20 W, which resulted in a deposition rate of 0.15 Å/s. After Sb2S3 deposition, the photoresist was removed by soaking the sample in a 1-Methyl-2-pyrrolidinone (NMP) solution placed in a 60 °C water bath for one hour. The sample was then rinsed with Acetone followed by Isopropyl Alcohol, before drying it with N$_2$ gas. An SEM image of a typical Sb$_2$S$_3$ strip is given in Fig. S4.

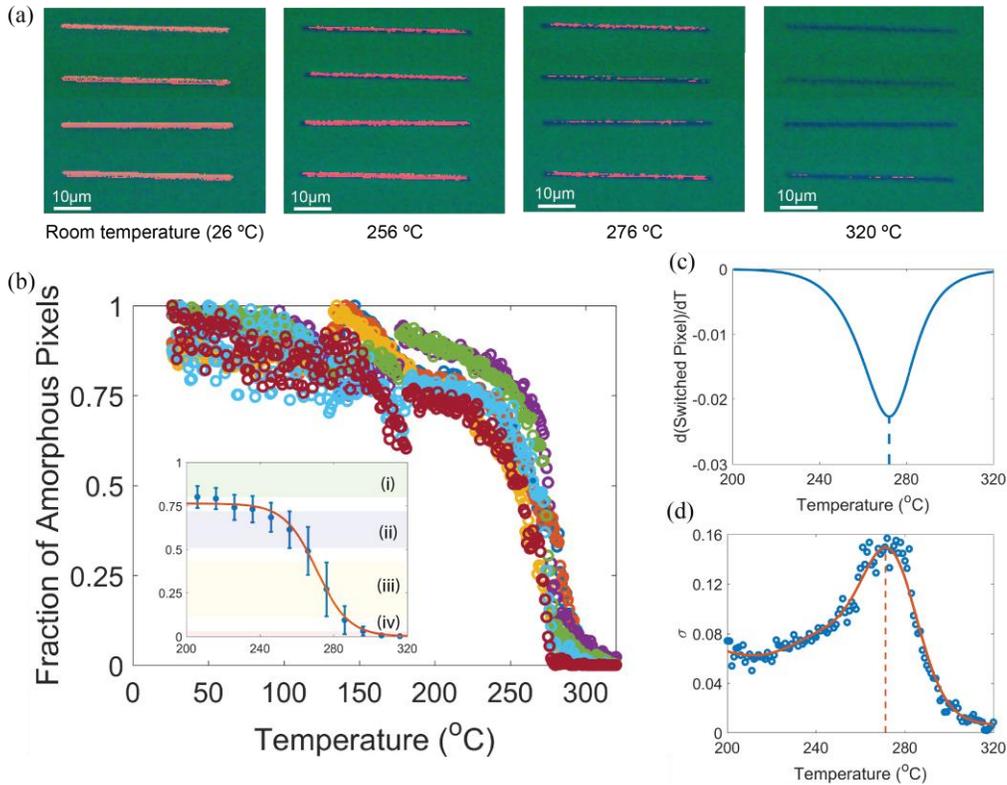

**Fig. 8.** Crystallization profile of Sb$_2$S$_3$ strips. (a) False color optical images of the Sb$_2$S$_3$ strip taken during the heating process. The color change represents phase transition. Pink and blue pixels represent amorphous and crystalline regions respectively. (b) Fraction of remaining amorphous pixels with respect to temperature of the seven Sb$_2$S$_3$ strips. Inset in (b) shows the corresponding average and standard deviation of switched pixel across the seven strips. The regions (i) to (iv) represent the four switching states that can be achieved. (c) Averaged differential of the seven switching curves in (b). The trough represents the average crystallization temperature. (d) Standard deviation of switched pixels with respect to temperature. The peak occurs at the crystallization temperature due to the stochastic nature of $Sb_2S_3$ crystallization.

The $Sb_2S_3$ strip crystallization behavior was studied by tracking the change in optical reflectivity normal to the strip's surface. The samples were heated to 320 °C using a heating rate of 5 °C/min in a Linkam microscope furnace (Linkam T95-HT) with an Ar gas flow rate of 4 SCCM. Microscope images of the strips under a ×10 objective lens were collected for every 1 °C rise in temperature, i.e at a rate of 5 images per minute. The strip increased in reflectivity during crystallization, which allowed the crystal growth to be monitored [31]. False color optical microscope images of the strips are shown in Fig. 8a where the amorphous and crystalline regions are shown as pink and blue, respectively. The strips tend to crystallize inwards from the edges. These results show that $Sb_2S_3$ strip crystallization is growth driven, as is also seen in continuous films [17]. Fig. 8b shows the fraction of the remaining amorphous pixels for seven different $Sb_2S_3$ strips as a function of temperature. For clarity and further analysis, the corresponding average and standard deviation of the fraction of switched pixels for every 10 °C rise in temperature is given in the inset of Fig. 8b. The average crystallization temperature was 270±5 °C, as shown in Fig. 8c.

The stochastic nature of $Sb_2S_3$ crystallization makes it challenging to reliably determine the transmissivity level. From Fig. 8d. we see that the standard deviation of switched pixels is largest at 270 °C, when the rate of crystallization is maximum. This is due to the strips crystallizing at slightly different temperatures. This variation can be attributed to the growth-driven crystallization kinetics. As nucleation is a stochastic process, the number of nucleation centers in the strip is random. This variability in crystallization was previously attributed to material defects [14], but is more likely due to the stochastic nature of $Sb_2S_3$ crystallization. Regardless of the origin, this variability will limit the number of switching levels because at a certain temperature, the crystallized length of the $Sb_2S_3$ strip varies.

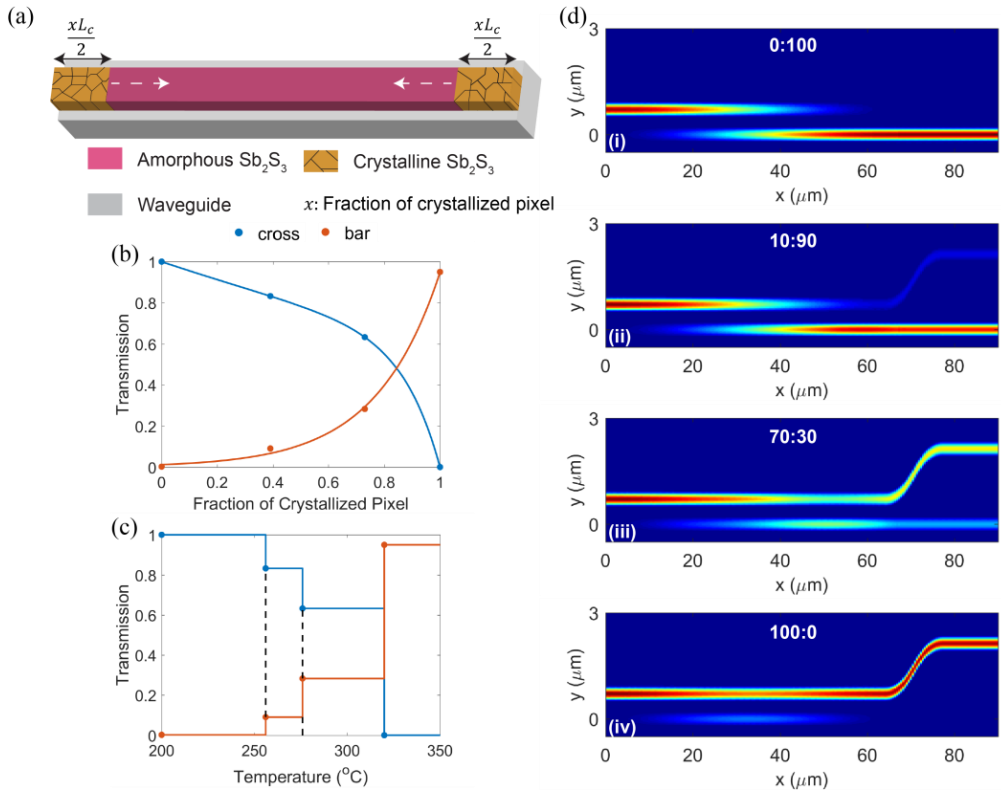

**Fig. 9.** 2-bit Sb$_2$S$_3$ directional coupler. (a) Partial crystallization model of the cross waveguide. The strip is sectioned into the amorphous and crystalline regions based on the experimentally measured fraction of crystallized pixels. Since the material tends to crystallize inwards from the end, the crystalline regions increase along the direction of the white arrows with increasing temperature. (b) Transmission through the cross and bar output port for the four crystalline lengths. (c) Operating temperature regions of the four coupling ratios (d) Power distribution and their corresponding coupling ratio of the Sb$_2$S$_3$ directional coupler base on the crystallization percentage. The coupling ratio is represented as cross: bar in each power field distribution plot. (i) to (iv) correspond to the colored regions in inset of Fig. 8b.

Temperature can be used to control the output power of the coupler to four coupling ratios, which in turn gives four switching states. Using the standard deviation of switched pixels at different temperatures, the extent of crystallization can be discerned into four statistically different levels over a 32 dB dynamic range. The dynamic range was approximated based on Figs. 5a and 5b. The switching states are presented as regions (i) to (iv) in the inset of Fig. 8b. The four levels can be programmed by heating the strips at their corresponding temperatures. To model the effect of partial crystallization on the waveguide transmission, the Sb$_2$S$_3$ strip in the coupler model was sectioned to the corresponding amorphous and crystalline regions. As the strip crystallizes inwards from the edges, we varied the length of the crystalline regions along the direction of the white arrows shown in Fig. 9a. The crystallized lengths of the two intermediate states are proportional to the fraction of crystallized pixels. 3D FDTD simulations were conducted to determine the directional coupler Cross and Bar transmissivity for each of the four discernable levels of crystallinity observed in the crystallization experiment in Fig 8b. Fig. 9b shows the corresponding transmission values of the bar and cross port for the four states at 1550 nm. The transmission values result in the coupling ratios of approximately: 100:0, 10:90, 70:30, and 0:100. Fig. 9c shows the operating temperature range to attain the four different coupling ratios and Fig. 9d illustrates how the power is distributed along the waveguide for each state. The corresponding coupling ratio at each state is also labelled in Fig.

9d. We see that by using the GCT method, temperature can be used to tune the Cross output port transmissivity to four distinct levels. Decreasing crystallization stochasticity will increase the number of transmissivity levels that can be reliably distinguished.

The GCT method must provide reliable and faster programming speeds when implemented on a PIC. This can be challenging as even our steady-state experimental setup showed variable crystallization behavior. To reliably program the devices at high speeds, the crystallization stochasticity should be minimized. One method involves retaining a portion of the crystal matrix in the amorphization process, as proposed by Zhang et al [39]. The stochastic nucleation process is avoided, and crystallization only involves crystal growth.

In addition to GCT, the two laser programming methods depicted in Fig. 10 can be adopted to further increase the $Sb_2S_3$ coupler bit-depth. In the first method, which is shown in Fig. 10a, the initial state of the $Sb_2S_3$ layer is crystalline and the *entire* length is partially amorphized to varying degrees to achieve the different switching states. Previous works have amorphized $Sb_2S_3$ thin films to varying degrees with a femtosecond laser [16, 40]. Moreover in reference [40] recyclability was demonstrated where the material can switch up to 7000 times depending on the extent of amorphization. The second method, which is shown in Fig. 10b, involves fully amorphizing regions of a crystalline $Sb_2S_3$ strip to implement the different switching states. The transmission level is then set by the length of the amorphous strip. Both processes use the fact that amorphisation is more deterministic than crystallization, and therefore the transmissivity can be controlled more accurately. Previous works have shown that $Sb_2S_3$ can be amorphized with ns and fs laser pulses [15, 16, 40]. Hence these methods may also be suitable for higher speed reprogramming.

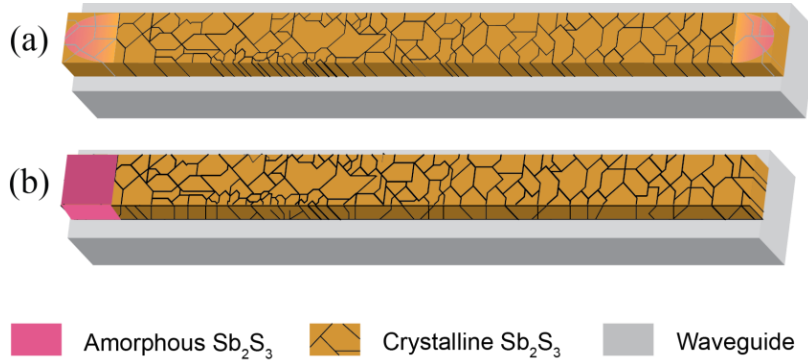

**Fig. 10.** Programming methods to minimize or avoid crystallization stochasticity. (a) Amorphizing the *entire* crystalline $Sb_2S_3$ strips to varying degrees to implement different switching states. This reduces stochasticity as nucleation of the material is avoided. (b) Amorphizing regions on the crystalline strip to implement the various switching states. The material is tuned by amorphizing different length of the strip, which is more deterministic that crystal nucleation and growth.

## 3. Conclusion

To conclude, we show that all PCM- tuned directional coupler models display low insertion losses (<1.5 dB) and low crosstalk (−20 dB to −40 dB) in both amorphous and crystalline states. $Sb_2S_3$ directional couplers have a better modelled overall performance than $Sb_2Se_3$, $Ge_2Sb_2Te_4S_1$, and $Ge_2Sb_2Te_5$ -based couplers. The $Sb_2S_3$ couplers show the lowest insertion loss in both the amorphous and crystalline state due to a lower extinction coefficient than the other PCMs. However, since the $Sb_2S_3$ - based coupler has the lowest $\Delta n$, a longer coupling length is needed to meet the $L_{c,amor} \approx 2 \cdot L_{c,crys}$ condition. The $\boldsymbol{k}$ value of Sb$_2$Se$_3$ is insufficiently low, which results in a high insertion loss in the crystalline state. These losses can be higher than the Telluride-based directional couplers, where the signal is not coupled into the neighboring waveguide in the crystalline state. The longer coupling length of Sb$_2$S$_3$ is actually an advantage when introducing multiple switching states with the GCT scheme because it is easier to control the fractional length of the waveguide crystallized for longer strips of PCM. For this reason and because Sb$_2$S$_3$ had the lowest losses, we also studied how it crystallizes when patterned as strips. This was to understand how Sb$_2$S$_3$ can be programmed into different coupling ratios. We found that four coupling ratios can be discerned across a dynamic range of 32 dB. These states are sufficient to implement 2-bit weights in a conceptual Sb$_2$S$_3$ programmable PIC quantized neural network, which could down-scale deep neural networks in space-limited platforms, such as mobile devices [41]. The Sb$_2$S$_3$ -coupler bit-depth can be further increased by retaining a portion of the crystal matrix to limit nucleation stochasticity or by programming the devices through amorphization.

**Funding.** This work was funded by the Agency for Science, Technology and Research (A*STAR) Advanced Manufacturing and Engineering (AME) grant (A18A7b0058); Czech Science Foundation, Grant number: 19-17997S; The Ministry of Education, Youth and Sports of the Czech Republic, Grant number: LM2018103.

**Acknowledgement.** The work was carried out under the auspices of the SUTD-MIT International Design Centre and the University of Pardubice. LL is grateful for his Ministry of Education Singapore PhD scholarship.

**Disclosures.** The authors declare no conflicts of interest.

**Data Availability.** Data underlying the results presented in this paper are not publicly available at this time but may be obtained from the authors upon reasonable request.

**Supplemental document.** See supplementary document for supporting content.